\documentclass[12pt]{article}
\usepackage{lipsum}
\usepackage{authblk}
\usepackage[top=2cm, bottom=2cm, left=2cm, right=2cm]{geometry}
\usepackage{fancyhdr}
\usepackage{xspace,amsmath,amsfonts,amsthm,amssymb,amsbsy}
\usepackage{graphicx,epstopdf}
\usepackage{bbm}
\usepackage[breaklinks=true]{hyperref}
\usepackage{color}

\usepackage{wrapfig}

\newcommand{\ket}[2] {| #1 \rangle_{#2}}
\newcommand{\bra}[2] {\langle #1 |_{#2}}
\newcommand{\dg}{^{\dagger}}
\newcommand{\ee}[1] {\mathrm{e}^{#1}}

\renewenvironment{abstract}{%
\hfill\begin{minipage}{0.95\textwidth}
\rule{\textwidth}{1pt}}
{\par\noindent\rule{\textwidth}{1pt}\end{minipage}}
\makeatletter
\renewcommand\@maketitle{%
\hfill
\begin{minipage}{0.95\textwidth}
\vskip 2em
\let\footnote\thanks 
{\LARGE \@title \par }
\vskip 1.5em
{\large \@author \par}
\end{minipage}
\vskip 1em \par
}
\makeatother
\begin{document}
%
\title{\textbf{Hong-Ou-Mandel Interference}}
\author{Agata M. Bra\'nczyk}
\affil{\small{\emph{Perimeter Institute for Theoretical Physics, Waterloo, Ontario, N2L 2Y5, Canada}}\\\small{abranczyk@pitp.ca}}
\maketitle

\begin{abstract}
This article is a detailed introduction to Hong-Ou-Mandel (HOM) interference, in which two photons interfere on a beamsplitter in a way that depends on the photons' distinguishability. We begin by considering distinguishability in the polarization degree of freedom. We then consider spectral distinguishability, and show explicitly how to calculate the HOM dip for three interesting cases: 1)  photons with arbitrary spectral distributions, 2)  spectrally entangled  photons, and 3) spectrally mixed photons. 
\end{abstract}

\section{Introduction}
When two indistinguishable photons interfere on a beam splitter, they behave in an interesting way. This effect is known as Hong-Ou-Mandel (HOM) interference, named after Chung Ki Hong, Zhe Yu Ou and Leonard Mandel, who experimentally verified the effect in 1987 \cite{Hong1987}. HOM interference shows up in many places, both in fundamental studies of quantum mechanics and in practical implementations of quantum technologies. At its heart, HOM interference is quite simple to understand. But it can also be very rich once different aspects of the incoming light are considered. 

This document contains a step-by-step account of  how these more interesting effects can be modelled. We begin,  in Section \ref{sec:basic},  with a basic model of two photons interfering on a beam splitter. We then consider the photons' polarization degree of freedom in Section \ref{sec:pol},  and show that the output state (after the beam splitter) is fundamentally different depending on the photons' relative polarizations. We then extend the model to account for the photons' spectro-temporal properties in Section \ref{sec:temp}, and show how to calculate the famous \emph{HOM dip}. As concrete examples, we examine three interesting cases: 1)  photons with arbitrary spectral distributions, 2)  spectrally entangled  photons, and 3) spectrally mixed photons.  

This document is intended to serve as a pedagogical guide; we therefore go into much more detail than in a typical research paper.

\section{A basic model of two-photon interference}\label{sec:basic}

\begin{wrapfigure}{r}{0.45\textwidth}
   \vspace{-0.5cm}
\begin{center}
    \includegraphics[width=0.3\columnwidth]{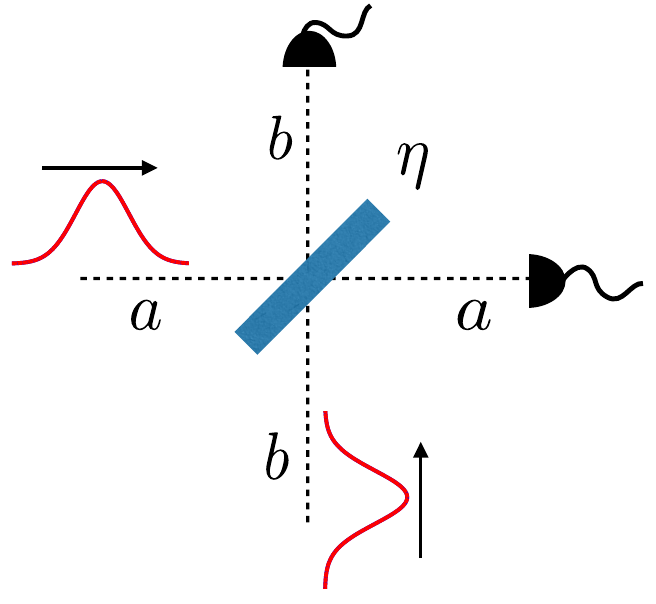}
    \end{center}
    \vspace{-0.5cm}
\caption{Two photons interfere on a beam splitter of reflectivity $\eta$. Upon exiting the beam splitter, the photons are detected. }
\label{fig:bsnew}
\end{wrapfigure}

Consider two photons incident on a beam splitter, as shown in Figure \ref{fig:bsnew}.  The combined two-photon state \emph{before} arriving at the beam splitter, i.e. the \emph{input} state, is:
\begin{align}
\ket{\psi_{\textrm{in}}}{ab}={}&\hat{a}\dg_{j}\hat{b}\dg_{k}\ket{0}{ab}=\ket{1;j}{a}\ket{1;k}{b}\,,
\end{align}
where $\hat{a}\dg_{j}$ and $\hat{b}\dg_{k}$ are bosonic creation operators in beam splitter modes, $a$ and $b$, respectively. In addition to being identified by their respective beam splitter modes, the photons can have other properties, labeled by $j$ and $k$, that determine how distinguishable they are. \\
Some examples of such additional properties are the photon's  polarization, spectral mode \cite{Eckstein2011,Sharapova2015}, temporal mode \cite{Ansari2017}, arrival time, or  transverse spatial mode \cite{Langford2004}. For the time being, we make no assumptions about the photons' level of distinguishability. 

\begin{figure}[h]
\begin{center}
\includegraphics[width=0.9\columnwidth]{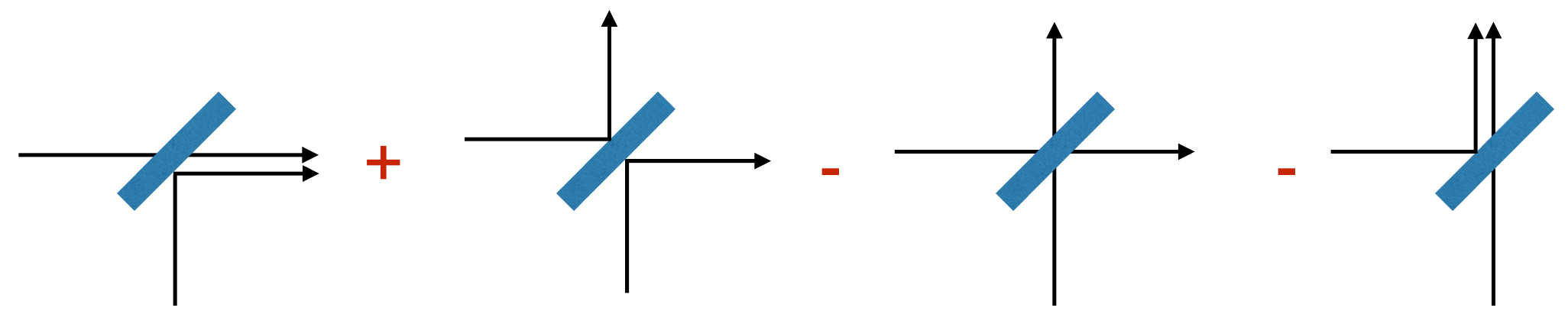}
\caption{Diagram showing four different ways for two photons to interact on a beam splitter. The signs correspond to signs in front of terms in Eq. (\ref{eq:4bs}) }
\label{fig:4bs}
\end{center}
\end{figure}

The evolution of a state as it interferes on a beam splitter with reflectivity $\eta$ can be modelled with a unitary $\hat{U}_{\textsc{bs}}$ \cite{Bachor2004}. The unitary acts on the creation operators as follows:
\begin{subequations}\label{eq:unitary}
\begin{align}
\hat{a}\dg\xrightarrow{\hat{U}_{\textsc{bs}}} {}&\sqrt{1-\eta}\hat{a}\dg+\sqrt{\eta}\hat{b}\dg\\
\hat{b}\dg\xrightarrow{\hat{U}_{\textsc{bs}}}{}&\sqrt{\eta}\hat{a}\dg-\sqrt{1-\eta}\hat{b}\dg\,.
\end{align}
\end{subequations}
The combined two-photon state \emph{after} exiting the beam splitter, i.e. the \emph{output} state, is then:
\begin{align}
\ket{\psi^{\mathrm{out}}}{ab}={}&\hat{U}_{\textsc{bs}}\ket{\psi^{\mathrm{in}}}{ab}\\
={}&\hat{U}_{\textsc{bs}}\left(\hat{a}\dg_{j}\hat{b}\dg_{k}\ket{0}{ab}\right)\\
={}&\left(\sqrt{1-\eta}\hat{a}\dg_j+\sqrt{\eta}\hat{b}\dg_j\right)\left(\sqrt{\eta}\hat{a}\dg_k-\sqrt{1-\eta}\hat{b}\dg_k\right)\ket{0}{ab}\\
={}&\left(\sqrt{\eta(1-\eta)}\hat{a}\dg_{j}\hat{a}\dg_{k}+\eta\hat{a}\dg_{k}\hat{b}\dg_{j}-(1-\eta)\hat{a}\dg_{j}\hat{b}\dg_{k}+\sqrt{\eta(1-\eta)}\hat{b}\dg_{j}\hat{b}\dg_{k}\right)\ket{0}{ab}\,.
\end{align}
In the case where $\eta=1/2$, the output state is 
\begin{align}\label{eq:4bs}
\ket{\psi^{\mathrm{out}}}{ab}={}&\frac{1}{2}\left(\hat{a}\dg_{j}\hat{a}\dg_{k}+\hat{a}\dg_{k}\hat{b}\dg_{j}-\hat{a}\dg_{j}\hat{b}\dg_{k}-\hat{b}\dg_{j}\hat{b}\dg_{k}\right)\ket{0}{ab}\,.
\end{align}
Figure \ref{fig:4bs} shows a diagram representing the four terms in Eq. (\ref{eq:4bs}). \\

In HOM interference, we are often interested in the \emph{coincidence probability}, that is, the probability of detecting one photon in each output port of the beam splitter. To compute this, we must  take into account the distinguishability of the input photons. In the next section, we consider distinguishability in the polarization degree of freedom.

\section{Polarization distinguishability}\label{sec:pol}

Just like classical light, an individual photon can be described as having horizontal ($H$) or vertical ($V$) polarization, or a superposition of the two ($\alpha H+\beta V$, where $\alpha$ and $\beta$ are complex numbers satisfying $|\alpha|^2+|\beta|^2=1$). In this section, we consider how the polarization of the input photons influences the HOM coincidence probability. 

\subsection{Distinguishable photons}

First  consider two photons with orthogonal polarizations, $H$ and $V$. Also assume that all other properties of the  photons (spectrum, arrival time, transverse spatial mode, etc.) are identical. Two photons with orthogonal polarizations are said to be \emph{distinguishable}. In this scenario, where $j=H$ and $k=V$, the output state is
\begin{align}
\ket{\psi^{\mathrm{out}}}{ab}={}&\frac{1}{2}\left(\hat{a}\dg_{H}\hat{a}\dg_{V}+\hat{a}\dg_{V}\hat{b}\dg_{H}-\hat{a}\dg_{H}\hat{b}\dg_{V}-\hat{b}\dg_{H}\hat{b}\dg_{V}\right)\ket{0}{ab}\\\label{eq:dist}
={}&\frac{1}{2}\left(\ket{1;H}{a}\ket{1;V}{a}+\ket{1;V}{a}\ket{1;H}{b}-\ket{1;H}{a}\ket{1;V}{b}-\ket{1;H}{b}\ket{1;V}{b}\right)\,.
\end{align}
The first term contains both photons in mode $a$, the second and third terms contain only one photon in each mode $a$ and $b$, and the fourth term contains both photons in mode $b$.

We can compute the coincidence probability from the probability amplitudes in front of terms with only one photon in each output port, i.e., the two middle terms in Eq. (\ref{eq:dist}). The coincidence probability is therefore $p=|1/2|^2+|{-}1/2|^2=1/2$.

\subsection{Indistinguishable photons}

Now consider that the two photons have the same polarization,  $j=k=H$. All other things being equal (spectrum, arrival time, transverse spatial mode, etc.), two photons with the same polarization are said to be \emph{in}distinguishable. In this scenario, the output state is

\begin{align}\label{eq:some}
\ket{\psi^{\mathrm{out}}}{ab}={}&\left(\hat{a}\dg_{H}\hat{a}\dg_{H}+\hat{a}\dg_{H}\hat{b}\dg_{H}-\hat{a}\dg_{H}\hat{b}\dg_{H}-\hat{b}\dg_{H}\hat{b}\dg_{H}\right)\ket{0}{ab}\\
={}&\left(\hat{a}\dg_{H}\hat{a}\dg_{H}-\hat{b}\dg_{H}\hat{b}\dg_{H}\right)\ket{0}{ab}\\\label{eq:indist}
={}&\frac{1}{\sqrt{2}}\left(\ket{2;H}{a}-\ket{2;H}{b}\right)\,.
\end{align}
Notice that the two middle terms in Eq. (\ref{eq:some}) cancel, but the state still comes out normalized because $(\hat{a}\dg)^{n}\ket{0}{}=\sqrt{n!}\ket{n}{}$. The first term in Eq. (\ref{eq:indist}) has both photons in mode $a$ and the second term has both photons in mode $b$, but there are no terms corresponding to one photon in each mode $a$ and $b$. Compare this with the output state for distinguishable photons in Eq. (\ref{eq:dist}). Incidentally, the state  in Eq. (\ref{eq:indist}) is sometimes called a two-photon N00N state, i.e., $\ket{N}{}\ket{0}{}+\ket{0}{}\ket{N}{}$ where $N=2$ \cite{Lee2002}.

Here, the coincidence probability of detecting one photon in each output mode is  $p=0$. So we see that when two indistinguishable photons interfere on a beam splitter of reflectivity $\eta=1/2$, the amplitudes for ``both transmitted'' and ``both reflected'' perfectly cancel out.

\section{Temporal distinguishability}\label{sec:temp}

\begin{figure}[h]
\begin{center}
\includegraphics[width=0.9\columnwidth]{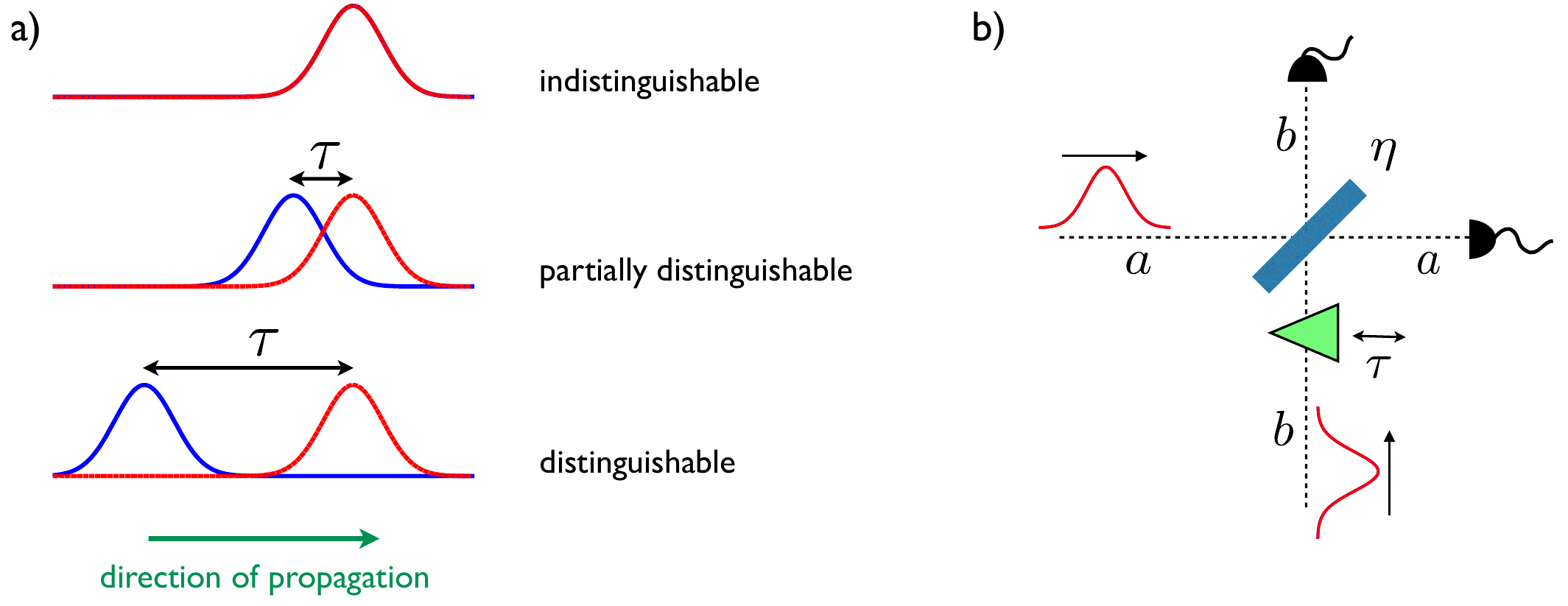}
\caption{a) Changing the time delay $\tau$ between two photons with  finite bandwidths changes how distinguishable they are. b) A time delay  in mode $b$ might be introduced by sending the photon in that mode  through a prism that introduces a phase shift. }
\label{fig:tune}
\end{center}
\end{figure}

Until now, we implicitly considered  photons with the same spectral and temporal properties (their details were thus not relevant to our analysis). We now extend our analysis to include the spectral profile of the photons, characterized by the spectral amplitude function $\phi(\omega)$, and the relative arrival times of the photons, parametrized by the time delay $\tau$.  By controlling the time delays between two such photons, it is possible to tune their level of distinguishability. This is shown schematically in Figure \ref{fig:tune}. 

\subsection{Photons with arbitrary spectra}\label{sec:arb}

The quantum state for a photon with spectral amplitude function  $\phi(\omega)$, in beam splitter mode $a$, is
\begin{align}
\ket{1;\phi}{a}=\int d\omega \phi(\omega)\hat{a}\dg(\omega)\ket{0}{a}\,,
\end{align}
where $\hat{a}\dg(\omega)$ represents a creation operator acting on a single frequency mode $\omega$. The state is normalized such that $\int d\omega|\phi(\omega)|^2=1$. \\

Now consider  two input photons with arbitrary  spectral amplitude functions $\phi$ and $\varphi$.  The two-photon input state is
\begin{align}\label{eq:onein}
\ket{\psi^{\mathrm{in}}}{ab}={}&\ket{1;\phi}{a}\ket{1;\varphi}{b}\\\label{eq:sep}
={}&\int d\omega_{1} \phi(\omega_{1})\hat{a}\dg(\omega_{1})\int d\omega_{2} \varphi(\omega_{2})\hat{b}\dg(\omega_{2})\ket{0}{ab}\,.
\end{align}

We are interested in how the coincidence probability changes as a function of the overlap between the photons. We thus introduce a time delay in, say, mode $b$. In practice, this might be done by sending the photon in mode $b$  through a prism that introduces a phase shift (see Figure \ref{fig:tune} b)), or perhaps by forcing it to take a longer path. The time delay has the following action on the creation operator:
\begin{align}\label{eq:td}
\hat{b}\dg(\omega)\rightarrow \hat{b}\dg(\omega)\ee{-i\omega \tau}\,.
\end{align}
The time-delayed state is
\begin{align}
\ket{\psi^{\mathrm{td}}}{ab}={}\int d\omega_{1} \phi(\omega_{1})\hat{a}\dg(\omega_{1})\int d\omega_{2} \varphi(\omega_{2})\hat{b}\dg(\omega_{2})\ee{-i\omega_{2} \tau}\ket{0}{ab}\,.
\end{align}

The beam splitter acts on each frequency mode independently, and we'll assume that the reflectivity is not frequency-dependent. The beam splitter unitary thus acts on the creation operators as follows:
\begin{align}\label{eq:bsomega}
\hat{a}\dg(\omega)\xrightarrow{\hat{U}_{\textsc{bs}}} {}& \sqrt{1-\eta}\hat{a}\dg(\omega)+\sqrt{\eta}\hat{b}\dg(\omega)\\
\hat{b}\dg(\omega)\xrightarrow{\hat{U}_{\textsc{bs}}} {}& \sqrt{\eta}\hat{a}\dg(\omega)-\sqrt{1-\eta}\hat{b}\dg(\omega)\,.
\end{align}
After passing through a beam splitter  with $\eta=1/2$,  the output state of the two photons is
\begin{align}
\ket{\psi^{\mathrm{out}}}{ab}={}&\hat{U}_{\textsc{bs}}\ket{\psi^{\mathrm{td}}}{ab}\\
={}&
\frac{1}{2}\int d\omega_{1} \phi(\omega_{1})\left(\hat{a}\dg(\omega_{1})+\hat{b}\dg(\omega_{1})\right)\int d\omega_{2} \varphi(\omega_{2})\left(\hat{a}\dg(\omega_{2})-\hat{b}\dg(\omega_{2})\right)\ee{-i\omega_{2} \tau}\ket{0}{ab}\\\label{eq:stateesr}
\begin{split}
={}&\frac{1}{2}\int d\omega_{1} \phi(\omega_{1})\int d\omega_{2} \varphi(\omega_{2})\ee{-i\omega_{2} \tau}\\
&\times\left(\hat{a}\dg(\omega_{1})\hat{a}\dg(\omega_{2})+\hat{a}\dg(\omega_{2})\hat{b}\dg(\omega_{1})-\hat{a}\dg(\omega_{1})\hat{b}\dg(\omega_{2})-\hat{b}\dg(\omega_{1})\hat{b}\dg(\omega_{2})\right)\ket{0}{ab}\,.
\end{split}
\end{align}

Earlier, when we considered photons of a single frequency,  it was simple to read off the coincidence probability from the state. Here, it is a bit more tricky so we should calculate it explicitly. We'll model each detector as having a flat frequency response. The projector describing detection in mode $a$ is given by
\begin{align}\label{eq:pa}
\hat{P}_{a}=\int d\omega\hat{a}\dg(\omega)\ket{0}{a}\bra{0}{a}\hat{a}(\omega)\,,
\end{align}
and the projector describing detection in mode $b$ is given by
\begin{align}\label{eq:pb}
\hat{P}_{b}=\int d\omega\hat{b}\dg(\omega)\ket{0}{b}\bra{0}{b}\hat{b}(\omega)\,.
\end{align}
The coincidence probability of detecting one photon in each mode is 
\begin{align}\label{eq:coin}
p={}&\mathrm{Tr}[\ket{\psi^{\mathrm{out}}}{ab}\bra{\psi^{\mathrm{out}}}{ab}\hat{P}_{a}\otimes\hat{P}_{b}]=\bra{\psi^{\mathrm{out}}}{ab}\hat{P}_{a}\otimes\hat{P}_{b}\ket{\psi^{\mathrm{out}}}{ab}\,.
\end{align}
For two photons with arbitrary spectral amplitude functions $\phi$ and $\varphi$, the coincidence probability is 
\begin{align}\label{eq:someeq}
\begin{split}
p_{\mathrm{arb}}={}&\Bigg[\frac{1}{2}\int d\omega_{1}\int d\omega_{2} \phi^*(\omega_{1}) \varphi^*(\omega_{2})\ee{i\omega_{2} \tau}\\
&\times\bra{0}{}\left(\hat{a}(\omega_{1})\hat{a}(\omega_{2})+\hat{a}(\omega_{2})\hat{b}(\omega_{1})-\hat{a}(\omega_{1})\hat{b}(\omega_{2})-\hat{b}(\omega_{1})\hat{b}(\omega_{2})\right)\Bigg]\\
&\times\Bigg[\int d\omega_{a}\hat{a}\dg(\omega_{a})\ket{0}{a}\bra{0}{a}\hat{a}(\omega_{a})\int d\omega_{b}\hat{b}\dg(\omega_{b})\ket{0}{b}\bra{0}{b}\hat{b}(\omega_{b})\Bigg]\\
&\times\Bigg[\frac{1}{2}\int d\omega'_{1} \int d\omega'_{2} \phi(\omega'_{1})\varphi(\omega'_{2})\ee{-i\omega'_{2} \tau}\\
&\times\left(\hat{a}\dg(\omega'_{1})\hat{a}\dg(\omega'_{2})+\hat{a}\dg(\omega'_{2})\hat{b}\dg(\omega'_{1})-\hat{a}\dg(\omega'_{1})\hat{b}\dg(\omega'_{2})-\hat{b}\dg(\omega'_{1})\hat{b}\dg(\omega'_{2})\right)\ket{0}{ab}\Bigg]\,,
\end{split}
\end{align}
where all we did so far was insert Eqs. (\ref{eq:stateesr}), (\ref{eq:pa}), and (\ref{eq:pb}) into Eq. (\ref{eq:coin}). Reshuffling some parts, we can write
\begin{align}
\begin{split}
p_{\mathrm{arb}}={}&\frac{1}{4}\int d\omega_{a}\int d\omega_{b}\int d\omega_{1} \int d\omega_{2}\int d\omega'_{1}\int d\omega'_{2}\phi^*(\omega_{1}) \varphi^*(\omega_{2}) \phi(\omega'_{1})\varphi(\omega'_{2})\ee{i(\omega_2-\omega'_{2}) \tau} \\
&\times\bra{0}{ab}\left(\hat{a}(\omega_{1})\hat{a}(\omega_{2})+\hat{a}(\omega_{2})\hat{b}(\omega_{1})-\hat{a}(\omega_{1})\hat{b}(\omega_{2})-\hat{b}(\omega_{1})\hat{b}(\omega_{2})\right)\hat{a}\dg(\omega_{a})\hat{b}\dg(\omega_{b})\ket{0}{ab}\\
&\times\bra{0}{ab}\hat{a}(\omega_{a})\hat{b}(\omega_{b})\left(\hat{a}\dg(\omega'_{1})\hat{a}\dg(\omega'_{2})+\hat{a}\dg(\omega'_{2})\hat{b}\dg(\omega'_{1})-\hat{a}\dg(\omega'_{1})\hat{b}\dg(\omega'_{2})-\hat{b}\dg(\omega'_{1})\hat{b}\dg(\omega'_{2})\right)\ket{0}{ab}\,.
\end{split}
\end{align}
Terms with an odd number of operators in one mode, e.g. $\bra{0}{ab}\hat{a}\hat{a}\hat{a}\dg\hat{b}\dg\ket{0}{ab}$ and $\bra{0}{ab}\hat{b}\hat{b}\hat{a}\dg\hat{b}\dg\ket{0}{ab}$, go to zero, while terms such as $\bra{0}{ab}\hat{a}\hat{b}\hat{a}\dg\hat{b}\dg\ket{0}{ab}$ give delta functions:
\begin{align}
\begin{split}
p_{\mathrm{arb}}={}&\frac{1}{4}\int d\omega_{a}\int d\omega_{b}\int d\omega_{1} \int d\omega_{2}\int d\omega'_{1}\int d\omega'_{2}\phi^*(\omega_{1}) \varphi^*(\omega_{2}) \phi(\omega'_{1})\varphi(\omega'_{2})\ee{i(\omega_2-\omega'_{2}) \tau} \\
&\times\left(\delta(\omega_2-\omega_a)\delta(\omega_1-\omega_b)-\delta(\omega_1-\omega_a)\delta(\omega_2-\omega_b)\right)\\
&\times\left(\delta(\omega'_2-\omega_a)\delta(\omega'_1-\omega_b)-\delta(\omega'_1-\omega_a)\delta(\omega'_2-\omega_b)\right)\,.
\end{split}
\end{align}
Using the delta functions to evaluate the integrals over $\omega_a$ and $\omega_b$ gives an expression with two terms:
\begin{align}
\begin{split}
p_{\mathrm{arb}}={}&\frac{1}{2}\int d\omega_{1} \int d\omega_{2}\int d\omega'_{1}\int d\omega'_{2}\phi^*(\omega_{1}) \varphi^*(\omega_{2}) \phi(\omega'_{1})\varphi(\omega'_{2})\ee{i(\omega_2-\omega'_{2}) \tau} \\
&\times\left(\delta(\omega_2-\omega'_2)\delta(\omega_1-\omega'_1)-\delta(\omega_1-\omega'_2)\delta(\omega_2-\omega'_1)\right)\,.
\end{split}
\end{align}
Using the  remaining delta functions to evaluate the integrals over $\omega'_1$ and $\omega'_2$, and taking advantage of the  normalization condition $\int d\omega|\phi(\omega)|^2=1$, gives
\begin{align}\label{eq:someth}
\begin{split}
p_{\mathrm{arb}}={}&\frac{1}{2}-\frac{1}{2}\int d\omega_{1}\phi^*(\omega_{1})\varphi(\omega_{1}) \ee{-i\omega_{1} \tau}\int d\omega_{2}\varphi^*(\omega_{2}) \phi(\omega_{2}) \ee{i\omega_2 \tau}\,.
\end{split}
\end{align}
If $\phi(\omega)=\varphi(\omega)$, this expression simplifies to 
\begin{align}\label{eq:someg}
\begin{split}
p_{\mathrm{arb}}={}&\frac{1}{2}-\frac{1}{2}\int d\omega_{1}|\phi(\omega_{1})|^2\ee{-i\omega_{1} \tau} \int d\omega_{2} |\phi(\omega_{2})|^2\ee{i\omega_2 \tau}\,.
\end{split}
\end{align}

\subsubsection{Example: Gaussian photons}

Consider two photons with Gaussian spectral amplitude functions, 
\begin{align}
\phi_i(\omega)=\frac{1}{( \pi )^{1/4}\sqrt{   \sigma_i }}e^{-\frac{\left(\omega -\bar\omega_i\right){}^2}{2 \sigma_i^2}}\,; ~~~~~(i=a,b)\,,
\end{align} 
where $\bar\omega_i$ is the central frequency of photon $i$, $\sigma_i$ defines its spectral width,   and the normalization was chosen such that $\int d\omega |\phi_i(\omega)|^2=1$. From Eq. (\ref{eq:someth}), the coincidence probability is
\begin{align}
\begin{split}
p_{\mathrm{arb,gauss}}={}&\frac{1}{2}-\frac{1}{2\pi  \sigma_a\sigma_b }\left(\int d\omega_{1} e^{-\frac{\left(\omega_1 -\bar\omega_a\right){}^2}{2 \sigma_a^2}}e^{-\frac{\left(\omega_1 -\bar\omega_b\right){}^2}{2 \sigma_b^2}}\ee{-i\omega_{1} \tau}\right)\left(\int d\omega_{2}e^{-\frac{\left(\omega_2 -\bar\omega_a\right){}^2}{2 \sigma_a^2}}e^{-\frac{\left(\omega_2 -\bar\omega_b\right){}^2}{2 \sigma_b^2}} \ee{i\omega_2 \tau}\right)\,.
\end{split}
\end{align}
The Fourier Transforms can be evaluated using your favourite method (mine is \emph{Mathematica}). The coincidence probability simplifies to
\begin{align}
\begin{split}
p_{\mathrm{arb,gauss}}={}&\frac{1}{2}-\frac{\sigma_a\sigma_b }{ (\sigma_a^2+\sigma_b^2 )}e^{-\frac{\sigma_a^2\sigma_b^2\tau^2+(\bar\omega_a-\bar\omega_b)^2}{\sigma_a^2+\sigma_b^2}}\,.
\end{split}
\end{align}
If the Gaussians are equal, that is $\phi_a=\phi_b$, we have
\begin{align}\label{eq:dsgds}
\begin{split}
p_{\mathrm{arb,gauss}}={}&\frac{1}{2}-\frac{1}{2}\ee{- \frac{\sigma_a ^2 \tau^2}{2}}\,.
\end{split}
\end{align}

\subsubsection{Example: Sinc-shaped photons}

Consider two photons with a sinc spectral amplitude function, 
\begin{align}
\varphi_i(\omega)=\sqrt{\frac{A_i}{\pi}}\mathrm{sinc}\left(A_i (\omega-\bar\omega_i)\right)\,; ~~~~~(i=a,b)\,,
\end{align} 
where $\bar\omega_i$ is the central frequency of photon $i$, $A_i^{-1}$ defines its spectral width, and the normalization was chosen such that $\int d\omega |\varphi_i(\omega)|^2=1$. From Eq. (\ref{eq:someth}), the coincidence probability is
\begin{align}
\begin{split}
p_{\mathrm{arb,sinc}}={}&\frac{1}{2}-\frac{A_aA_b}{2 \pi^2 }\left(\int d\omega_{1} \mathrm{sinc}\left(A_a(\omega_1-\bar\omega_1)\right)\mathrm{sinc}\left(A_b(\omega_1-\bar\omega_1)\right)\ee{-i\omega_{1} \tau}\right)\\
&\times\left(\int d\omega_{2} \mathrm{sinc}\left(A_a(\omega_2-\bar\omega_2)\right)\mathrm{sinc}\left(A_b (\omega_2-\bar\omega_2)\right) \ee{i\omega_2 \tau}\right)\,.
\end{split}
\end{align}
Computing the Fourier Transforms, in the case where $\varphi_a=\varphi_b$, the coincidence probability simplifies to
\begin{align}\label{eq:dsgdsdfd}
\begin{split}
p_{\mathrm{arb,sinc}}={}&\frac{1}{2}-\frac{1}{8A_a^2} \left(  |\tau| -\left|\frac{ \tau}{2} -A_a\right| -\left|\frac{  \tau}{2} +A_a\right|\right)^2\,.
\end{split}
\end{align}

\begin{figure}[h]
\begin{center}
SEPARABLE PHOTONS\\
\includegraphics[width=0.5\columnwidth]{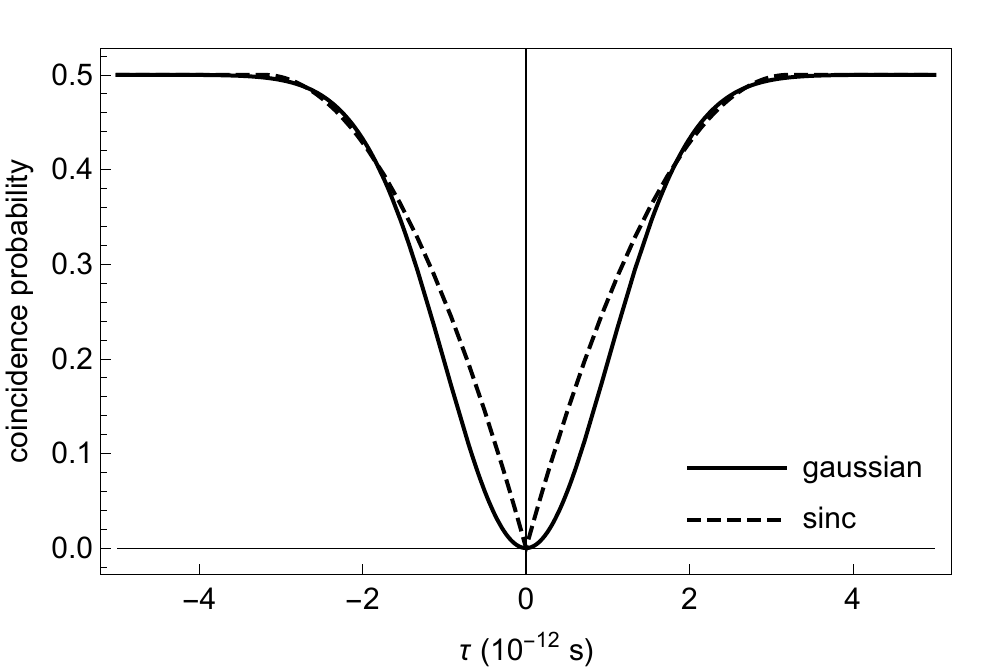}
\caption{The coincidence probability for separable photons, as a function of time delay $\tau$.  The solid line is  $p_{\mathrm{arb,gauss}}$ defined in Eq.  (\ref{eq:dsgds}) for $\sigma_a=10^{-12}$ rad/s, and the dashed line is $p_{\mathrm{arb,sinc}}$ defined in Eq. (\ref{eq:dsgdsdfd}) for $A_a=1/\sigma_a\sqrt{2\gamma}$ where $\sigma_a=10^{-12}$ rad/s and $\gamma=0.193$. This value of $\gamma$ ensures that the Gaussian and sinc spectral amplitude functions have the same widths, for comparison. }
\label{fig:arbhom}
\end{center}
\end{figure}

Figure \ref{fig:arbhom} compares the coincidence probabilities, Eqs. (\ref{eq:dsgds}) and (\ref{eq:dsgdsdfd}), for photons with Gaussian and sinc spectral amplitude functions respectively.

\subsection{Spectrally entangled photons}

In the previous section, we considered two photons with arbitrary, but separable, spectral amplitude profiles. More generally, however, the spectral amplitudes of two photons can be correlated--- that is, the photons can be entangled in the spectral degree of freedom. The nature of this spectral entanglement is captured by the joint spectral amplitude (JSA) function $f(\omega_1,\omega_2)$. The quantum state of two spectrally entangled photons is
\begin{align}\label{eq:entangled}
\ket{\psi^{\mathrm{in}}}{ab}={}&\int d\omega_1\int d\omega_2f(\omega_1,\omega_2)\hat{a}\dg(\omega_1)\hat{b}\dg(\omega_2)\ket{0}{ab}\,.
\end{align}
Notice that for $f(\omega_{1},\omega_{2})=\phi(\omega_{1})\varphi(\omega_{2})$, this state reduces to the separable state in Eq. (\ref{eq:sep}). Entangled photons can be sent onto a beam splitter in exactly the same way as separable photons, but the way they interfere will also depend on the nature of their entanglement.  

As before, to calculate the HOM dip, we begin by  introducing a time delay $\tau$ in mode $b$, by applying the transformation in Eq. (\ref{eq:td}). We then model how the photons interact via the beam splitter by applying the beam splitter unitary in Eq. (\ref{eq:bsomega}) to the time-delayed state. We finally calculate the coincidence probability, as defined in Eq. (\ref{eq:coin}), using  projectors, defined in Eqs. (\ref{eq:pa}) and  (\ref{eq:pb}), that describe detection in modes $a$  and $b$. The steps are identical to  those in Section \ref{sec:arb}, and yield the coincidence probability:
\begin{align}\label{eq:dggdgd}
\begin{split}
p_{\mathrm{ent}}={}&\frac{1}{2}- \frac{1}{2}\int d\omega_{1} \int d\omega_{2}f^*(\omega_{1},\omega_{2}) f(\omega_{2},\omega_{1})\ee{i(\omega_2-\omega_{1}) \tau}\,.
\end{split}
\end{align}
In fact, this is equivalent to the coincidence probability in Eq. (\ref{eq:someth}) when $f(\omega_{1},\omega_{2})=\phi(\omega_{1})\varphi(\omega_{2})$. In general, however, $f(\omega_{1},\omega_{2})$ will not take such a nice form, and the integrals in Eq. (\ref{eq:dggdgd}) will need to be evaluated numerically. 

It can sometimes be useful to express the JSA in terms of its' Schmidt decomposition,
\begin{align}
f(\omega_1,\omega_2)={}&\sum_{k}u_k \phi_k(\omega_1)\varphi_k(\omega_2)\,,
\end{align}
where the Schmidt modes $\phi_k(\omega)$ and $\varphi_k(\omega)$ each form a discrete basis of complex orthonormal functions ($\int d\omega \phi^*_k(\omega)\phi_{k'}(\omega)=\int d\omega \varphi^*_k(\omega)\varphi_{k'}(\omega)=\delta_{kk'}$), and the Schmidt coefficients ${u}_{k}$ are real and satisfy $\sum_{k}{u}_{k}^2=1$ if $f(\omega_{1},\omega_{2})$ is normalized. The coincidence probability can then be expressed in terms of the Schmidt coefficients and Schmidt modes as
\begin{align}
p_{\mathrm{ent}}={}&\frac{1}{2}-\frac{1}{2}\sum_{k,k'}u_ku_{k'}\int d\omega_{1}  \phi^*_k(\omega_1)\varphi_{k'}(\omega_1)\ee{-i\omega_1\tau} \int d\omega_{2}\varphi^*_k(\omega_2)\phi_{k'}(\omega_2)\ee{i\omega_2\tau}\,.
\end{align}

\subsubsection{Example: SPDC pumped by a pulsed pump laser}

The joint spectral amplitude for photons generated via spontaneous parametric downconversion is 
\begin{align}
f(\omega_1,\omega_2)\propto \Phi(\omega_{1},\omega_{2})\alpha(\omega_{1}+\omega_{2})\,,
\end{align}
such that $\int d\omega_1 d\omega_2|f(\omega_1,\omega_2)|^2=1$, where $\Phi(\omega_{1},\omega_{2})$ is known as the phase-matching function and $\alpha(\omega_1+\omega_2)$ is the pump amplitude function \cite{Grice1997}. 

A typical SPDC crystal of length $L$ generates a phase-matching function with a sinc profile:
\begin{align}
\Phi_{\mathrm{sinc}}(\omega_1,\omega_2)=\mathrm{sinc}\left(\frac{\Delta k(\omega_1,\omega_2)L}{2}\right)\,,
\end{align}
where $\Delta k(\omega_1,\omega_2)=k_p(\omega_1+\omega_2)-k_1(\omega_1)+k_2(\omega_2)$, and $k_i(\omega)$ are the wave numbers associated with the respective fields. 

\begin{figure}[t]
~~~~~~a)\hspace{7.5cm}b)
\vspace{-1.3cm}
\begin{center}
\includegraphics[width=0.35\columnwidth]{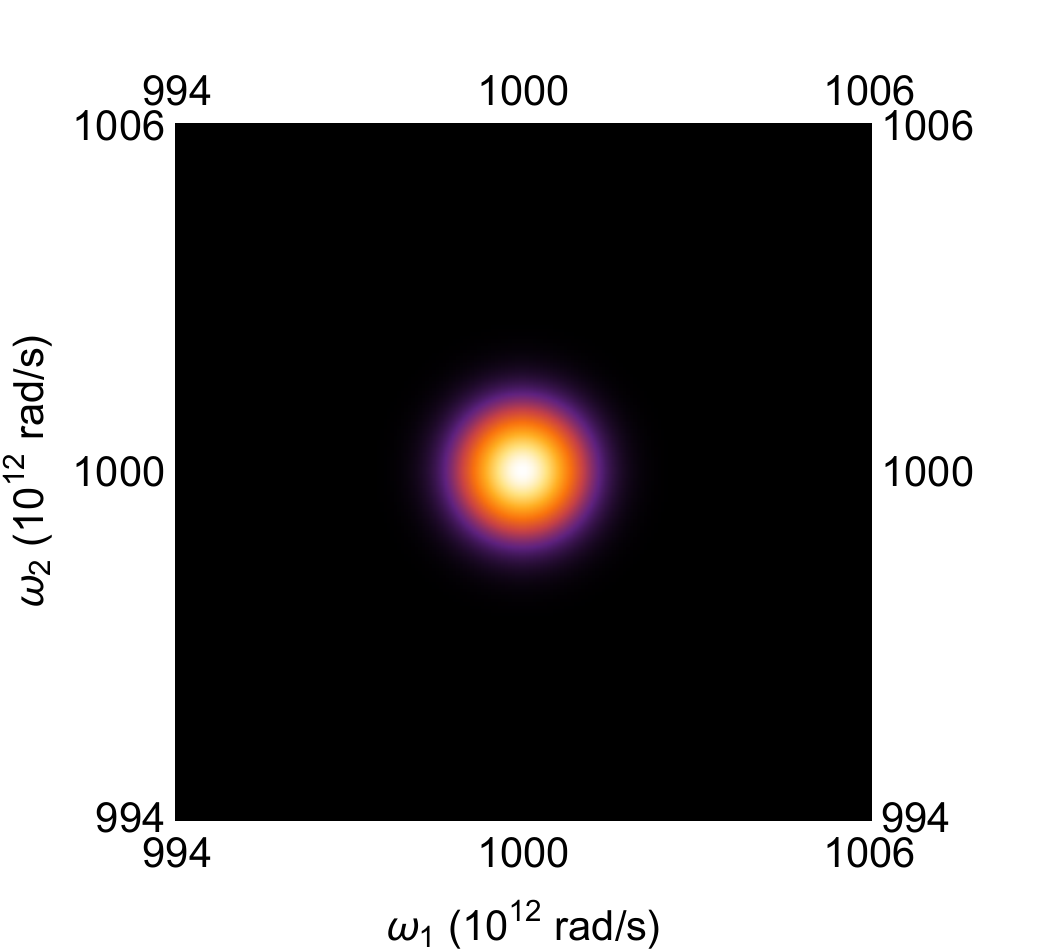}\hspace{1cm}\includegraphics[width=0.35\columnwidth]{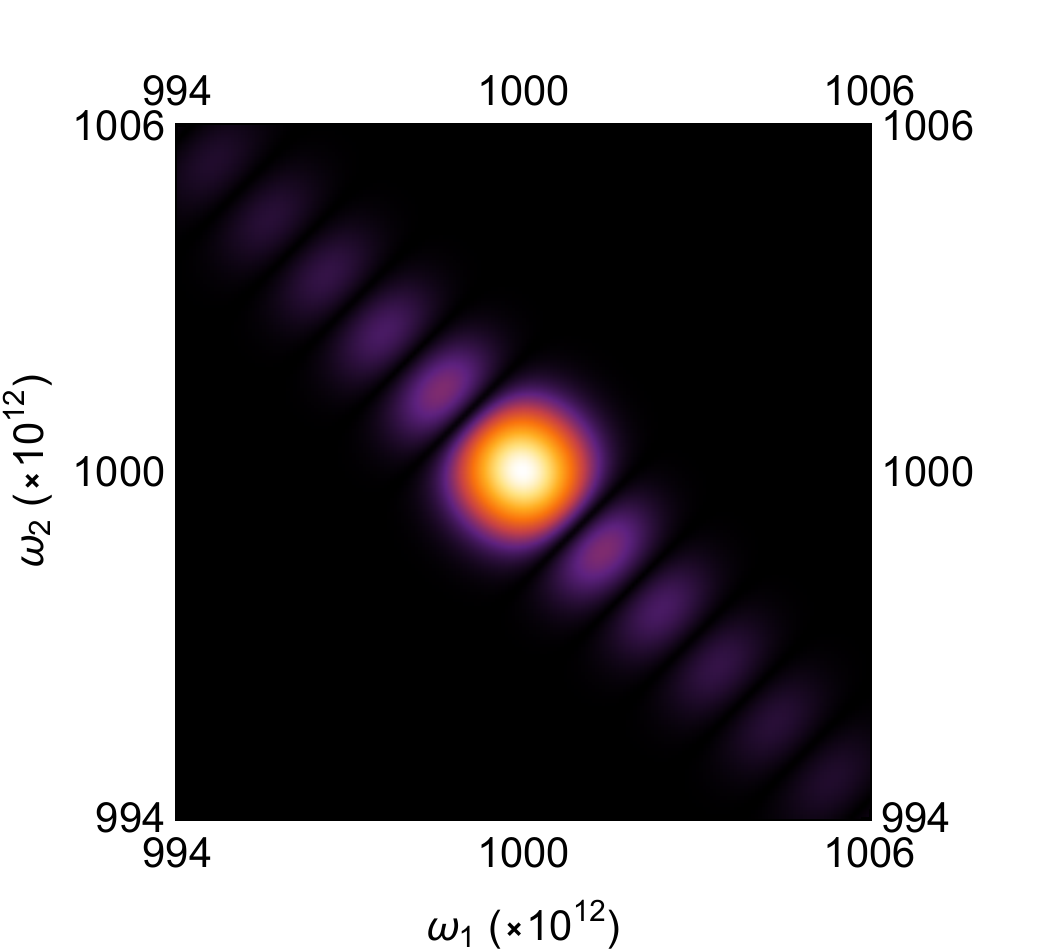}
\caption{ Joint spectral amplitudes: a) $|f_{\mathrm{gauss}}(\omega_1,\omega_2)|$, and b) $|f_{\mathrm{sinc}}(\omega_1,\omega_2)|$, with $A=1/\sigma\sqrt{2\gamma}$, $B=-A$,  $C=0$, $\gamma=0.193$, $\sigma=10^{12}$ rad/s, and $\bar\omega=10^{15}$ rad/s. }
\label{fig:jsa}
\end{center}
\end{figure}

To simplify calculations, we can Taylor expand $\Delta k(\omega_1,\omega_2)$ to first order:
\begin{align}
\Delta k(\omega_1,\omega_2)=k_{1,0}+k_{2,0}-k_{p,0}+k'_{1}(\omega_1-\bar\omega)+k'_{2}(\omega_2-\bar\omega)-k'_{p}(\omega_1+\omega_2-2\bar\omega)\,,
\end{align}
where $k_{p,0}=k_p(2\bar\omega)$,  $k_{1/2,0}=k_{1/2}(\bar\omega)$, $k_p'=\partial_p(\omega)/\partial\omega|_{\omega=2\bar\omega}$, and $k_{1/2}'=\partial_{1/2}(\omega)/\partial\omega|_{\omega=\bar\omega}$.   This approximation is valid in many regimes. We can then write
\begin{align}\label{eq:sinccc}
\Phi_{\mathrm{sinc}}(\omega_1,\omega_2)=\mathrm{sinc}\left(A\omega_1+B\omega_2-C\right)\,,
\end{align}
where 
\begin{align}
A={}&\frac{L}{2}(k'_{1}-k'_{p})\\
B={}&\frac{L}{2}(k'_{2}-k'_{p})\\
C={}&\frac{L}{2}(k_{1,0}+k_{2,0}-k_{p,0}+(k'_{1}+k'_{2}-2k'_{p})\bar\omega)\,.
\end{align}
For comparison, it is useful to define a Gaussian phase-matching function of the same width:
\begin{align}\label{eq:gausss}
\Phi_{\mathrm{gauss}}(\omega_1,\omega_2)=e^{-\gamma\left(A\omega_1+B\omega_2-C\right)^2}\,,
\end{align}
where the parameter $\gamma=0.193$ ensures that the Gaussian and sinc functions have the same widths. Gaussian phase-matching functions were originally used in the literature to simplify calculations. But, more recently, methods have been developed to generate them in practice \cite{Branczyk2011, Dixon2013, Dosseva2016, Tambasco2016, Graffitti2017}. In combination with the right set of parameters (A, B, C, and $\sigma$), Gaussian phase-matching functions make it possible to  generate  separable joint spectral amplitudes via SPDC (see Fig. \ref{fig:jsa} a))

For a pulsed pump laser, it is common to assume a Gaussian pump amplitude function:
\begin{align}
\alpha(\omega)=e^{-\frac{(\omega-\bar\omega)^2}{2\sigma^2}}\,,
\end{align}
where $\bar\omega$ is the central frequency of the pump, and $\sigma$ defines the spectral width. 

We define the corresponding joint spectral amplitudes:
\begin{align}\label{eq:fgauss}
f_{\mathrm{gauss}}(\omega_1,\omega_2)\propto {}&\Phi_{\mathrm{gauss}}(\omega_{1},\omega_{2})\alpha(\omega_{1}+\omega_{2})\\\label{eq:fsinc}
f_{\mathrm{sinc}}(\omega_1,\omega_2)\propto {}&\Phi_{\mathrm{sinc}}(\omega_{1},\omega_{2})\alpha(\omega_{1}+\omega_{2})\,,
\end{align}
which are plotted in Figure \ref{fig:jsa}.

\begin{figure}[h!]
\begin{center}
SPDC PAIR (PULSED)\\
\includegraphics[width=0.5\columnwidth]{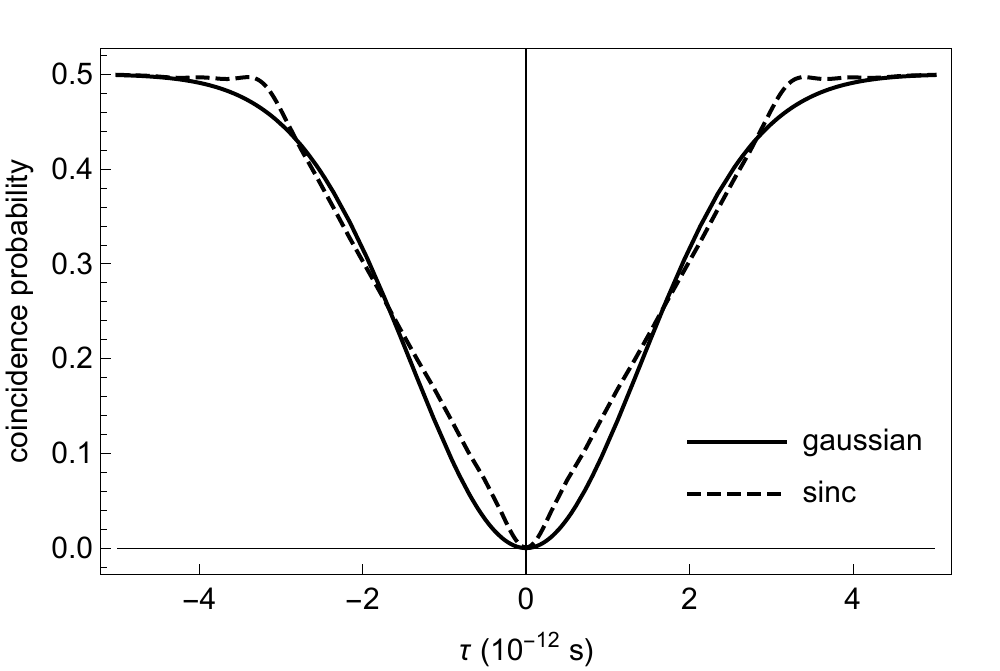}
\caption{The coincidence probability for photons from an SPDC source pumped by a pulsed laser, Eq. (\ref{eq:dggdgd}), as a function of time delay $\tau$. The solid line is for $f_{\mathrm{gauss}}(\omega_1,\omega_2)$ defined in Eq.  (\ref{eq:fgauss}), and the dashed line is for $f_{\mathrm{sinc}}(\omega_1,\omega_2)$ defined in Eq.  (\ref{eq:fsinc}). All parameters are the same as for Figure \ref{fig:jsa}.}
\label{fig:ent}
\end{center}
\end{figure}

Figure \ref{fig:ent} compares the coincidence probabilities for photons from an SPDC source pumped by a pulsed laser, with Gaussian and sinc phase matching functions functions respectively. 


\subsubsection{Example: SPDC pumped by a CW pump laser}

For a nonlinear source pumped by a continuous wave (CW) laser at frequency $2\bar\omega$, the joint spectral amplitude takes the form 
\begin{align}
f(\omega_1,\omega_2)\propto \Phi(\omega_1,\omega_2)\delta(\omega_1+\omega_2-2\bar\omega)\,,
\end{align}
such that $\int d\omega_1 d\omega_2|f(\omega_1,\omega_2)|^2=1$. The coincidence probability in Eq. (\ref{eq:dggdgd}) then simplifies to 
\begin{align}\label{eq:cwprob}
p_{\mathrm{cw}}={}&\frac{1}{2}-\frac{1}{2}\int d\omega g^*(-\omega)g(\omega)\ee{i2\omega \tau} \,,
\end{align}
where  $g(\omega)\propto\Phi(\bar\omega-\omega,\bar\omega+\omega)$, such that $\int d\omega|g(\omega)|^2=1$. 

\begin{figure}[h]
\begin{center}
SPDC PAIR (CW)\\
\includegraphics[width=0.5\columnwidth]{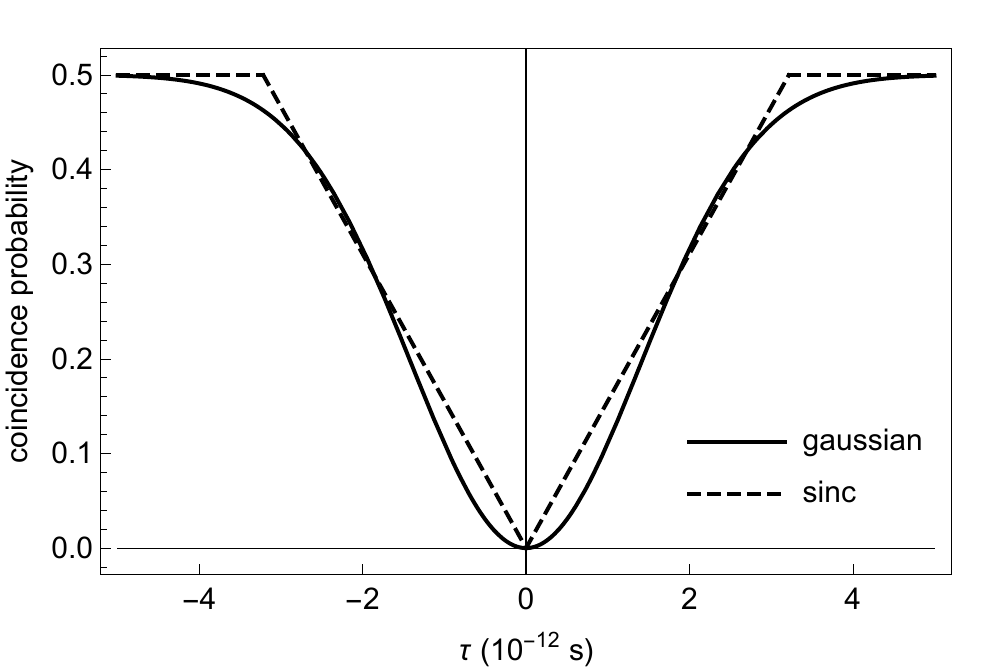}
\caption{The coincidence probability for entangled photons from an SPDC source pumped by a CW laser, Eq. (\ref{eq:cwprob}), as a function of time delay $\tau$. The solid line is for  $\Phi_{\mathrm{gauss}}(\omega_1,\omega_2)$ defined in Eq.  (\ref{eq:gausss}), and the dashed line is  for $\Phi_{\mathrm{sinc}}(\omega_1,\omega_2)$ defined in Eq.  (\ref{eq:sinccc}). All parameters are the same as for Figure \ref{fig:jsa}.}
\label{fig:entcw}
\end{center}
\end{figure}

Figure \ref{fig:entcw} compares the coincidence probabilities for photons from an SPDC source pumped by a CW laser, with Gaussian and sinc phase matching functions functions respectively. 

\subsection{Spectrally mixed states}

In the previous sections, we considered single photons and photon pairs in spectrally pure states. It is possible, however, for a photon to be in a mixture of different spectral amplitude functions  $\phi_k(\omega)$. This can be represented by the density matrix 
\begin{align}
\rho_{\phi}={}&\sum_{k}q_{k}\ket{1;\phi_{k}}{a}\bra{1;\phi_{k}}{a}\,,
\end{align} 
where 
\begin{align}
\ket{1;\phi}{a}={}&\int d\omega \phi(\omega)\hat{a}\dg(\omega)\ket{0}{a}\,,
\end{align}
is the quantum state for a single photon in a spectral mode defined by the spectral amplitude function $\phi(\omega)$. The density matrix $\rho_{\phi}$ can be realized in one of two ways: 1) as a probabilistic preparation on the pure state $\ket{1;\phi_{k}}{}$ with probability $q_{k}$, or 2) as the reduced density matrix of a spectrally entangled two-photon state, such as Eq. (\ref{eq:entangled}) (see Appendix \ref{sec:red} for details). 

Spectrally mixed photons can be sent onto a beam splitter in exactly the same way as separable or entangled photons. The density operator for a two-photon input state can be written as 
\begin{align}
\rho^{\mathrm{in}}={}&\rho_{\phi}\otimes\rho_{\varphi}\\
={}&\sum_{k}q_{k}\ket{1;\phi_{k}}{a}\bra{1;\phi_{k}}{a}\otimes \sum_{k}q'_{k'}\ket{1;\varphi_{k'}}{b}\bra{1;\varphi_{k'}}{b}\\
={}&\sum_{kk'}q_{k}q'_{k'}\left(\ket{1;\phi_{k}}{a}\ket{1;\varphi_{k'}}{b}\right)\left(\bra{1;\phi_{k}}{a}\bra{1;\varphi_{k'}}{b}\right)\,.
\end{align}

As before, the next step is to introduce a time delay $\tau$ in mode $b$, by applying the transformation in Eq. (\ref{eq:td}), and then model how the photons interact via the beam splitter by applying the beam splitter unitary in Eq. (\ref{eq:bsomega}) to the time-delayed state. But notice that the state $\ket{1;\phi_{k}}{a}\ket{1;\varphi_{k'}}{b}$ is just the state of two photons with arbitrary spectral amplitude functions $\phi_{k}$ and $\varphi_{k'}$ that we saw in Eq. (\ref{eq:onein}) in Section \ref{sec:arb}. Due to the linearity of quantum mechanics, we can simply use the result from Section \ref{sec:arb} to write the output density operator 
\begin{align}
\rho^{\mathrm{out}}={}&\sum_{kk'}q_{k}q'_{k'}\ket{\psi_{kk'}^{\mathrm{out}}}{ab}\bra{\psi_{kk'}^{\mathrm{out}}}{ab}\,,
\end{align}
where
\begin{align}
\begin{split}
\ket{\psi_{kk'}^{\mathrm{out}}}{ab}={}&\frac{1}{2}\int d\omega_{1} \phi_k(\omega_{1})\int d\omega_{2} \varphi_{k'}(\omega_{2})\ee{-i\omega_{2} \tau}\\
&\times\left(\hat{a}\dg(\omega_{1})\hat{a}\dg(\omega_{2})+\hat{a}\dg(\omega_{2})\hat{b}\dg(\omega_{1})-\hat{a}\dg(\omega_{1})\hat{b}\dg(\omega_{2})-\hat{b}\dg(\omega_{1})\hat{b}\dg(\omega_{2})\right)\ket{0}{ab}\,.
\end{split}
\end{align}
This is equivalent to Eq. (\ref{eq:stateesr}) for $\phi(\omega)=\phi_{k}(\omega)$ and $\varphi(\omega)=\varphi_{k'}(\omega)$. The coincidence probability of getting one photon in each mode is 
\begin{align}
p_{\mathrm{mix}}={}&\mathrm{Tr}[\rho^{\mathrm{out}}\hat{P}_{a}\otimes\hat{P}_{b}]\\
={}&\sum_{kk'}q_{k}q'_{k'}\bra{\psi_{kk'}^{\mathrm{out}}}{ab}\hat{P}_{a}\otimes\hat{P}_{b}\ket{\psi_{kk'}^{\mathrm{out}}}{ab}\,,
\end{align}
where $\bra{\psi_{kk'}^{\mathrm{out}}}{ab}\hat{P}_{a}\otimes\hat{P}_{b}\ket{\psi_{kk'}^{\mathrm{out}}}{ab}$ is the coincidence probability, defined in Eq. (\ref{eq:coin}), for two photons with arbitrary spectral amplitude functions $\phi_{k}(\omega)$ and $\varphi_{k'}(\omega)$. We can therefore replace $\bra{\psi_{kk'}^{\mathrm{out}}}{ab}\hat{P}_{a}\otimes\hat{P}_{b}\ket{\psi_{kk'}^{\mathrm{out}}}{ab}$ with Eq. (\ref{eq:someth}), for $\phi(\omega)=\phi_{k}(\omega)$ and $\varphi(\omega)=\varphi_{k'}(\omega)$, to get 
\begin{align}\label{eq:mixp}
\begin{split}
p_{\mathrm{mix}}={}&\frac{1}{2}-\frac{1}{2}\sum_{kk'}q_{k}q'_{k'}\int d\omega_{1}\phi_{k}^*(\omega_{1})\varphi_{k'}(\omega_{1}) \ee{-i\omega_{1} \tau}\int d\omega_{2}\varphi_{k'}^*(\omega_{2}) \phi_{k}(\omega_{2}) \ee{i\omega_2 \tau}\,.
\end{split}
\end{align}

\subsubsection{Example: Independent SPDC sources}

\begin{figure}[h]
\begin{center}
MIXED STATES FROM INDEPENDENT SPDC SOURCES\\
\includegraphics[width=0.5\columnwidth]{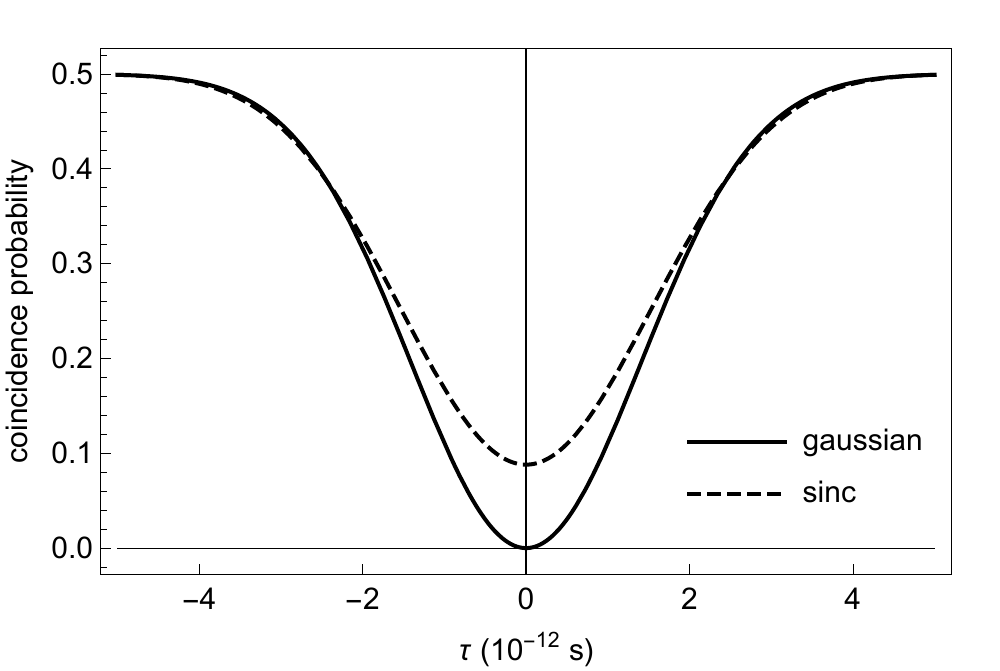}
\caption{The coincidence probability for photons from two independent SPDC sources pumped by pulsed lasers, Eq. (\ref{eq:pmix}), as a function of time delay $\tau$. The solid line is for $f(\omega_1,\omega_2)=h(\omega_1,\omega_2)=f_{\mathrm{gauss}}(\omega_1,\omega_2)$, where $f_{\mathrm{gauss}}(\omega_1,\omega_2)$ is defined in Eq.  (\ref{eq:fgauss}), and the dashed line is for $f(\omega_1,\omega_2)=h(\omega_1,\omega_2)=f_{\mathrm{sinc}}(\omega_1,\omega_2)$, where  $f_{\mathrm{sinc}}(\omega_1,\omega_2)$ is defined in Eq.  (\ref{eq:fsinc}). All parameters are the same as for Figure \ref{fig:jsa}.}
\label{fig:mix}
\end{center}
\end{figure}

Consider two independent SPDC sources that generate the entangled states:
\begin{align}
\ket{\psi_1}{ab}={}&\int d\omega_1\int d\omega_2f(\omega_1,\omega_2)\hat{a}\dg(\omega_1)\hat{b}\dg(\omega_2)\ket{0}{ab}\\
\ket{\psi_2}{cd}={}&\int d\omega_1\int d\omega_2h(\omega_1,\omega_2)\hat{c}\dg(\omega_1)\hat{d}\dg(\omega_2)\ket{0}{cd}\,.
\end{align}
To model a HOM experiment between photons in modes $a$ and $c$, we first compute the reduced density operators for those modes (see Appendix \ref{sec:red}):
\begin{align}
\rho_{\phi}={}&\mathrm{tr}_{b}\left[\ket{\psi_1}{ab}\bra{\psi_1}{ab}\right]=\sum_{k}u_1^2\ket{1;\phi_k}{a}\bra{1;\phi_{k}}{a}\\
\rho_{\varphi}={}&\mathrm{tr}_{d}\left[\ket{\psi_2}{cd}\bra{\psi_2}{cd}\right]=\sum_{k}v_1^2\ket{1;\varphi_k}{c}\bra{1;\varphi_{k}}{c}\,,
\end{align}
where $\phi_k$ and $\varphi_k$ are defined in terms of the Schmidt decompositions of the joint spectral amplitudes:
\begin{align}
f(\omega_1,\omega_2)={}&\sum_{k}u_k \phi_k(\omega_1)\phi'_k(\omega_2)\\
h(\omega_1,\omega_2)={}&\sum_{k}v_k \varphi_k(\omega_1)\varphi'_k(\omega_2)\,.
\end{align}
The coincidence probability, Eq. (\ref{eq:mixp}), becomes
\begin{align}\label{eq:pmix}
\begin{split}
p_{\mathrm{mix}}={}&\frac{1}{2}-\frac{1}{2}\sum_{kk'}u^2_{k}v^2_{k'}\int d\omega_{1}\phi_{k}^*(\omega_{1})\varphi_{k'}(\omega_{1}) \ee{-i\omega_{1} \tau}\int d\omega_{2}\varphi_{k'}^*(\omega_{2}) \phi_{k}(\omega_{2}) \ee{i\omega_2 \tau}\,.
\end{split}
\end{align}
Figure \ref{fig:mix} compares the coincidence probabilities for photons from two independent SPDC sources pumped by pulsed lasers, with Gaussian and sinc phase matching functions functions respectively. 

\subsubsection{Purity and Visibility}

In the special case of mixed, identical, and separable photons, there is a nice relationship between the visibility of the HOM dip and the purity of the input photons. 

The visibility of the HOM dip is given by
\begin{align}
V=\frac{p_{\mathrm{max}}-p_{\mathrm{min}}}{p_{\mathrm{max}}}\,,
\end{align}
where
\begin{align}
p_{\mathrm{max}}={}&\lim_{\tau\rightarrow\infty}p_{\mathrm{mix}}=\frac{1}{2}\\
p_{\mathrm{min}}={}&\lim_{\tau\rightarrow0}p_{\mathrm{mix}}=\frac{1}{2}-\frac{1}{2}\sum_{kk'}q_{k}q'_{k'}\int d\omega_{1}\phi_{k}^*(\omega_{1})\varphi_{k'}(\omega_{1})\int d\omega_{2}\varphi_{k'}^*(\omega_{2}) \phi_{k}(\omega_{2}) \,.
\end{align}

Given two photons with the same mixed density matrix ($\varphi_{k}=\phi_k$), 
\begin{align}
p_{\mathrm{min}}={}&\frac{1}{2}-\frac{1}{2}\sum_{kk'}q_{k}q_{k'}\int d\omega_{1}\phi_{k}^*(\omega_{1})\phi_{k'}(\omega_{1})\int d\omega_{2}\phi_{k'}^*(\omega_{2}) \phi_{k}(\omega_{2})\\
={}&\frac{1}{2}-\frac{1}{2}\sum_{kk'}q_{k}q_{k'}\int d\omega_{1}\phi_{k}^*(\omega_{1})\phi_{k'}(\omega_{1})\delta_{kk'}\\
={}&\frac{1}{2}-\frac{1}{2}\sum_{k}q^2_{k}\,.
\end{align}

In this case, the visibility is
\begin{align}
V={}&\frac{\frac{1}{2}-(\frac{1}{2}-\frac{1}{2}\sum_{k}q^2_{k})}{\frac{1}{2}}\\
={}&\sum_{k}q^2_{k}\,.
\end{align}

The visibility is then equal to the purity of each photon
\begin{align}
\mathrm{purity}={}&\mathrm{tr}\left[\rho_{\phi}^2\right]\\
={}&\mathrm{tr}\left[\sum_{k}q_{k}\ket{1;\phi_{k}}{a}\bra{1;\phi_{k}}{a}\sum_{k'}q_{k'}\ket{1;\phi_{k'}}{a}\bra{1;\phi_{k'}}{a}\right]\\
={}&\sum_{kk'}q_{k}q_{k'}\delta_{kk'}\\
={}&\sum_{k}q^2_{k}\,.
\end{align}

\section{(Not so) Final words}

In this document, we introduced Hong-Ou-Mandel interference in terms of the photons' distinguishability in the polarization degree of freedom. We then examined distinguishability as a function of the temporal overlap between the photons, and introduced the HOM dip. We saw how the HOM dip depends on the spectral properties of the input photons; in particular, how the HOM dip depends on the photons' spectral amplitude functions, as well as their entanglement with each other and with other photons. We also saw some examples relevant to photons generated via  spontaneous parametric downconversion (SPDC).

The observations in these notes are not new, and similar results are scattered throughout the literature (e.g. \cite{Fearn1989,Legero2003,Ou2006,Ou2007,Cosme2008}). It was my aim, however, to provide a self-contained pedagogical resource for students and researchers who want to see how these calculations are done explicitly. 

I intend for these notes to be a work-in-progress. In future versions, I'd like to include the effects of spectral filtering \cite{Branczyk2010}, multi-photon states \cite{Ou2006,Cosme2008}, and interference on multi-port beam splitters \cite{Tichy2010,Tan2013}. I would also like to include examples relevant to quantum-dot sources \cite{Dousse2010,Reimer2012,Versteegh2014}, which not only have different spectral amplitude functions, but also unique features such as time jitter in the emitted photon. If there are other examples that you would like included in further versions of these notes, please contact me.

\section{Acknowledgements}

An earlier informal version of these notes has been floating around the internet since 2012. I'd like to thank Hubert de Guise for motivating me to write them in the first place. Over the years, I've been pleasantly surprised by the number of people that read them and contacted me to say that they found them useful. This made me think it would be worthwhile to make a more formal version for the arXiv, with (hopefully) fewer typos and slip-ups. I'd also like to thank Francesco Graffitti and Morgan Mastrovich for making useful suggestions that I've incorporated in this version. 

\appendix

\section{Reduced density matrix}\label{sec:red}
The density matrix
\begin{align}\label{eq:dfdsgd}
\rho_{\phi}={}&\sum_{k}q_{k}\ket{1;\phi_{k}}{}\bra{1;\phi_{k}}{}\,,
\end{align} 
where
\begin{align}
\ket{1;\phi}{}={}&\int d\omega \phi(\omega)\hat{a}\dg(\omega)\ket{0}{}\,,
\end{align}
can be realized as the reduced density matrix of a spectrally entangled two-photon state. In other words, by preparing a spectrally entangled state such as the one introduced in Eq. (\ref{eq:entangled}),
\begin{align}
\ket{\psi^{\mathrm{in}}}{ab}={}&\int d\omega_1\int d\omega_2f(\omega_1,\omega_2)\hat{a}\dg(\omega_1)\hat{b}\dg(\omega_2)\ket{0}{ab}\,,
\end{align}
and discarding one of the photons. Mathematically, this is represented by ``tracing out'' the discarded mode using the partial trace operation. To perform the partial trace, we first make use of the  Schmidt decomposition to write
\begin{align}
\ket{\psi^{\mathrm{in}}}{ab}={}&\sum_{k}u_k \ket{1;\phi_k}{a}\ket{1;\varphi_k}{b}\,.
\end{align}
The reduced density matrix for system $a$ is
\begin{align}
\rho={}&\mathrm{tr}_{b}\left[\ket{\psi^{\mathrm{in}}}{ab}\bra{\psi^{\mathrm{in}}}{}\right]\\
={}&\sum_{kk'}u_k u_{k'} \mathrm{tr}_{b}\left[\ket{1;\phi_k}{a}\ket{1;\varphi_k}{b}\bra{1;\phi_{k'}}{a}\bra{1;\varphi_{k'}}{b}\right]\\
={}&\sum_{kk'}u_k u_{k'} \ket{1;\phi_k}{a}\bra{1;\phi_{k'}}{a}\mathrm{tr}\left[\ket{1;\varphi_k}{b}\bra{1;\varphi_{k'}}{b}\right]\\
={}&\sum_{kk'}u_k u_{k'} \ket{1;\phi_k}{a}\bra{1;\phi_{k'}}{a}\delta_{kk'}\\
={}&\sum_{k}u^2\ket{1;\phi_k}{a}\bra{1;\phi_{k}}{a}\,,
\end{align}
which has the same form as Eq. (\ref{eq:dfdsgd}) for $q_{k}=u^2_k$.

\end{document}